\documentclass[11pt,a4paper]{article}
\usepackage[hyperref]{acl2021}
\usepackage{times}
\usepackage{latexsym}
\usepackage{graphicx}
\usepackage{amsmath}
\usepackage{amssymb} 
\usepackage{multirow}
\usepackage{array}
\usepackage{booktabs} 
\usepackage{hyperref}
\usepackage{multirow}
\usepackage{subcaption}
\usepackage{caption}
\aclfinalcopy

\usepackage{microtype}

\title{Improving Document Representations by Generating Pseudo Query Embeddings for Dense Retrieval}

\author{Hongyin Tang$^{1,2,}$\thanks{\quad These authors contributed equally. This work was done when the first author was an intern at Meituan.}, Xingwu Sun$^{3,}$\footnotemark[\value{footnote}], Beihong Jin$^{1,2,}$\thanks{\quad Corresponding author.},  Jingang Wang$^{3}$, Fuzheng Zhang$^{3}$, Wei Wu$^{3}$\\
  $^1$State Key Laboratory of Computer Science \\
  Institute of Software, Chinese Academy of Sciences \\
  $^2$University of Chinese Academy of Sciences, Beijing China \\
  $^3$Meituan \\
  {\tt tanghongyin14@otcaix.iscas.ac.cn} \\
  \tt Beihong@iscas.ac.cn \\
  \tt \{sunxingwu, wangjingang02, zhangfuzheng, wuwei30\}@meituan.com
  }

\date{}

\begin{document}
\maketitle
\begin{abstract}
Recently, the retrieval models based on dense representations have been gradually applied in the first stage of the document retrieval tasks, showing better performance than traditional sparse vector space models. To obtain high efficiency, the basic structure of these models is Bi-encoder in most cases. However, this simple structure may cause serious information loss during the encoding of documents since the queries are agnostic. To address this problem, we design a method to mimic the queries on each of the documents by an iterative clustering process and represent the documents by multiple pseudo queries (i.e., the cluster centroids). To boost the retrieval process using approximate nearest neighbor search library, we also optimize the matching function with a two-step score calculation procedure. Experimental results on several popular ranking and QA datasets show that our model can achieve state-of-the-art results.
\end{abstract}

\section{Introduction}
Given a query and a collection of documents, the document retrieval task is to rank the documents based on their relevance with the query. To retrieve the target documents efficiently, most existing work adopts a two-stage fashion which retrieves a subset of candidate documents from the whole corpus by a recall model and then re-rank them by a sophisticated ranking model. In the first stage, many approaches use traditional information retrieval methods including BM25 based on sparse bag-of-word representation. Since the recall of the first-stage model determines the upper bound of the ranking quality, there is lots of work focusing on improving the recall performance\cite{DeepCT2019,doc2query,doct5query}.


In contrast to the sparse representations, dense representations encoding semantic information can enhance the retrieval performance by overcoming the limitations like term mismatching. They are usually produced by neural encoders whose parameters are learnable. Recently, inspired by the great success of pre-trained language models like BERT/RoBERTa\cite{BERT2018,roberta2019} in NLP applications, the dense passage retriever is proposed which encodes the documents by fine-tuning the huge language models \cite{DPR2020} and achieves state-of-the-art results benefiting from their powerful contextual semantic representative ability. 

Following the typical fine-tuning paradigm on many NLP tasks\cite{BERT2018}, a BERT encoder usually takes the concatenation of the query and document text as input and performs a full self-attention across the input tokens. Such architecture is called Cross-encoder \cite{poly2019}. Although it can achieve better performance than other architectures, it is infeasible in the recall stage since it needs to recompute the representation of each document in the corpus once a new query is provided. In contrast, Bi-encoder\cite{poly2019} encodes the queries and documents separately and computes the matching scores between their dense representations. Since the documents in the corpus keep unchanged most of the time, the representation of the documents can be stored in advance for future use. With the help of Approximate Nearest Neighbor (ANN) search approaches\cite{faiss17}, the retrieval process can be further boosted.


Although gaining retrieval efficiency, Bi-encoder sacrifices retrieval accuracy comparing to the Cross-encoder. To enrich the representations of the documents produced by Bi-encoder, some researchers extend the original Bi-encoder by employing more delicate structures like later-interaction\cite{Colbert2020}, Poly-Encoder\cite{poly2019}, multi-embedding model\cite{MEBERT2020}. Increasing a little computational overhead, these models can gain much improvement of the encoding quality while remaining the fast retrieval characteristic of Bi-encoder. 

Similar to these models, we focus on improving the effectiveness of Bi-encoder. In this work, we think that the limitation of the Bi-encoder origins from the agnostic nature of query when encoding the documents independently, i.e., the encoder cannot know what query could be potentially answered by the input document. As it is very common that a document with hundreds of tokens contains several distinct topics, some important semantic information might be easily missed or biased by each other without knowing the query. 

To alleviate the query agnostic problem, we propose a novel approach that mimics multiple potential queries corresponding to the input document and we call them ``pseudo query embeddings''. Ideally, each of the pseudo query embeddings corresponds to a semantic salient fragment in the document which is similar to a semantic cluster of the document. Thus, we implement the process by a clustering algorithm (i.e., K-means in this work) and regard the cluster centroids as the pseudo query embeddings. We generate and store all of the embeddings in an offline manner, thereby not only improving the encoding quality but also remaining the online computation unchanged.  During the inference, the multiple pseudo query embeddings should be first aggregated through a softmax function and then the relevance score with the query embedding is computed. Unfortunately, directly applying softmax aggregation is not supported in the existing ANN search library. Thus, we first filter some documents in which all of the embeddings have low relevance scores and then perform the whole aggregation and score function using the filtered embeddings.


Our main contributions can be summarized as follows:
\begin{itemize}
    \item We propose a novel approach to represent the document with multiple pseudo query embeddings which are generated by a clustering process.
    \item We modify the embedding aggregation during the inference in order to directly utilize the off-the-shelf ANN search library.
    \item We conduct experiments on several popular IR and OpenQA datasets. Experimental results show that our approach achieves state-of-the-art retrieval performance while still remaining efficient computation. An in-depth analysis on gradients shows how the cluster centroids improve the performance.
\end{itemize}

\section{Related Work}

In this section, we will review the existing work related with the first-stage retrieval. According to the representations of text, the first stage retrieval approaches can be classified into two categories. One is based on the high-dimensional sparse representation and the other is based on the low-dimensional continuous representation. Traditional sparse vector space models weight the terms by their frequency information. In last few years, some researchers intend to weight the document and query terms adaptively by a neural network which could leverage some semantical information \cite{DehghaniZSKC17,Zheng15}. Recently, a trend of leveraging the deep pre-trained language models to weight or augment the document/query terms is emerged. DeepCT\cite{DeepCT2019} uses BERT to learn the term importance and weight all of the terms. DocT5query\cite{doct5query} augments the document with possible query terms which are generated by a sequence-to-sequence model.

In contrast, the dense retrieval approaches map the text to continuous vectors which are mostly generated by neural networks. Models like DSSM\cite{dssm2013},CLSM\cite{Shen14}, DESM\cite{Mitra16} encode the query and document using their n-gram features or word embeddings independently and then compute their similarities. Recently, the dense retrieval approaches also tend to make use of the pre-trained language models. Sentence-BERT\cite{sentencebert} is a typical Bi-encoder model which encodes the text using BERT and calculates the similarity scores by the combination of several basic operations. Inspired by the interaction-based neural re-rankers, Khattab and Zaharia(\citeyear{Colbert2020}) propose a later-interaction mechanism. Later on, some variants\cite{mose2020,DiPair2020} are proposed. Xiong et al.(\citeyear{ance2020}) identify that the negative samples during training may not be representative, lowering the training difficulty. Therefore, they propose a model to construct hard negative samples dynamically during training. 

Comparing to existing work, our work serves the first stage of document retrieval and presents a new method to generate document representations which borrows the clustering technique to generate pseudo query embeddings from documents.

\section{Dense Document Retrieval}
In this section, we introduce the original Bi-encoder architecture and several existing variants. Then, we present our model in detail and describe the similarities and differences between our model and those Bi-encoder variants.
\subsection{Preliminaries}

\begin{figure*}
    \centering
    \includegraphics[width=\linewidth]{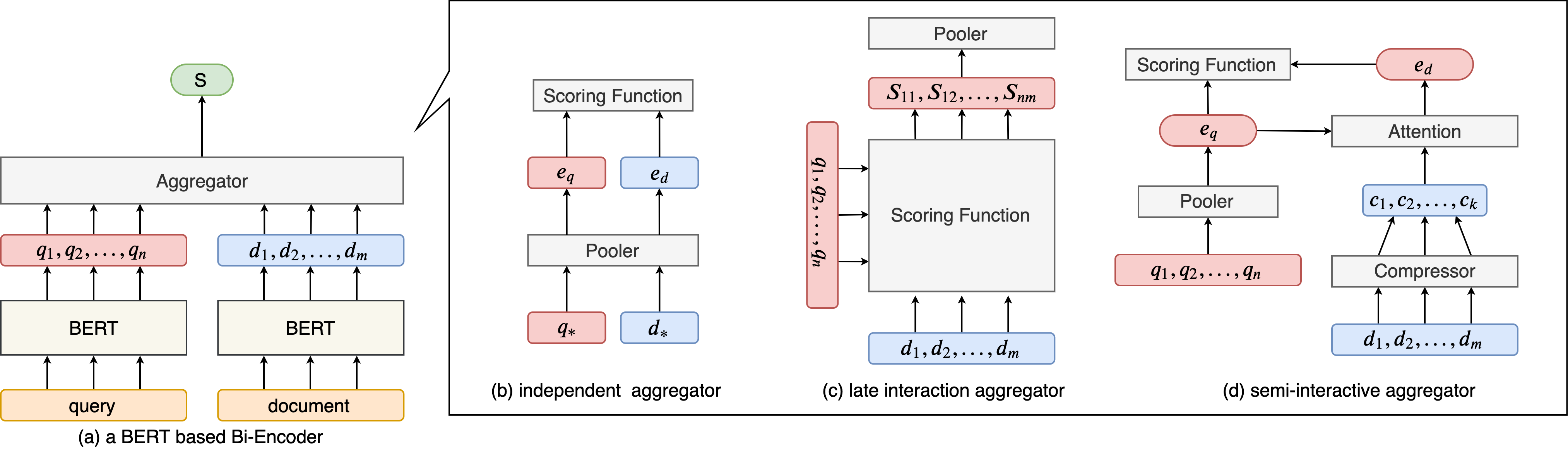}
    \caption{Bi-encoder and different aggregators}
    \label{fig:1}
\end{figure*}

\vspace*{1\baselineskip} 
\noindent{\bf Independent Aggregator} We start with a Bi-encoder using BERT as its backbone neural network as shown in Figure \ref{fig:1}. (a). Given a query with $n$ tokens and a document with $m$ tokens, a typical Bi-encoder encodes the query and the document separately, producing query token embeddings $\{q_i\}_{i=1}^n\in\mathbb{R}^{n\times h}$ and document token embeddings $\{d_i\}_{i=1}^m\in\mathbb{R}^{m\times h}$ which are the hidden states of the last layer in most cases. Next, a module is needed to compute the matching score by aggregating the generated query and document  representations. We call it ``Aggregator'' in the following sections. The simplest aggregator is the independent aggregator shown in Figure \ref{fig:1} (b). This aggregator uses a pooler to reduce the query and document token embeddings to fixed-length embeddings $e_q$ and $e_d$ respectively and then calculates the score by dot product/Euclidean distance between them. For example, Karpukhin et al.(\citeyear{DPR2020}) directly adopt the embedding of the \texttt{[CLS]} token. RepBERT\cite{RepBERT2020} leverages the mean value of the encoded embeddings. Although efficient to compute, compressing $m$ or $n$ ($m,n>>1$) embeddings to one may lose information. 

\noindent{\bf Late Interaction Aggregator} Col-BERT model\cite{Colbert2020} employs a late interaction paradigm to reduce  the loss of information. As shown in Figure \ref{fig:1} (c), the model preserves all of the document token embeddings $\{d_i\}_{i=1}^m$ in the cache until a new query is given. It then computes token-wise matching scores using all of the document and query embeddings. The final matching score is generated by pooling the $m\times n$ scores. This model preserves document semantics as much as possible and leaves the full query-document interaction during the inference. Experimental results show that Col-BERT is highly effective, improving the accuracy in a large margin. However, the time complexity of the score computation arises from constant $O(1)$ to quadratic $O(mn)$. Meanwhile, Lin et al.(\citeyear{survey2020}) point out that the storage space occupation also arises rapidly along with the length of documents since Col-BERT needs to store all of the embeddings. 

\vspace*{1\baselineskip} 
{\bf Semi-interactive Aggregator} Figure \ref{fig:1}(d) shows another kind of aggregator which compresses the document token embeddings to a constant number $k$ much smaller than the document length $m$ ($k<<m$). Since there are multiple but not all document token embeddings participating the interaction with query, we call the aggregator as a ``semi-interactive aggregator''. \cite{poly2019, MEBERT2020} adopt this aggregator in their model. Specifically, Poly-Encoder(learnt-k) \cite{poly2019} model employs $k$ learnable code-vectors as the parameters and attend them with all of the document token embeddings $\{d_i\}_{i=1}^m$, representing global features of the document. Besides, Poly-Encoder(first-k) \cite{poly2019} and ME-BERT\cite{MEBERT2020} both adopt the first $k$ document token embeddings as the compressed document representation. Obviously, the semi-interactive aggregator further makes time/space complexity and accuracy trade-offs over the independent aggregator and late interaction aggregator. However, there still exists some problem when applying current compressing strategies in the document retrieval task, which we would point out in the next section.

\subsection{Our Method}
The primary limitation of Bi-encoder is that we cannot know which part of the document would be asked during the encoding process. Preserving multiple semantic representations has been proved effective in the variants of Bi-encoder. However, existing models are still not perfect, leading to expensive computation or underfit problem. In this work, we intend to improve the semantic representations by mimicing the real matching process using the documents alone, generating a constant number of ``pseudo query embeddings''. In this way, the model can preserve self-adaptive document embeddings representing different semantics. Actually, the whole procedure is analogous to the steps of the K-means clustering algorithm and the cluster centroids are treated as the pseudo query embeddings. In the following, we will interpret the approach using the K-means algorithm in detail.


Firstly, following the semi-interactive aggregator, we feed the document tokens into BERT and use the last layer hidden states as the document token embeddings $\{d_i\}_{i=1}^m$. Next, we perform K-means algorithm on these token embeddings.

The K-means algorithm mainly contains two iterative steps: assignment step and update step. These two steps are performed alternatively until the convergence condition is satisfied. The assignment step can be expressed by the following equation.

\begin{equation}
\label{eq1}
\begin{split}
s_i^t &= \operatorname*{argmin}_j\|d_i-c_j^t\|^2 \\
    i\in&\{1,...,m\},j\in\{1,...,k\} 
\end{split}
\end{equation}
where $c_j^t$ is the $j$-th cluster centroid (we assume there are up to $k$ clusters) when the algorithm is executing at the $t$-th time step. $s_i^t$ represents the nearest cluster to the $i$-th embedding $d_i$ considering the Euclidean distance. After the assignment step, the algorithm updates each of the cluster centroid according to the cluster assignment of each embedding. The update step is shown as Eq. \ref{eq2}.

\begin{equation}
    c_j^{t+1} = \frac{1}{\sum_{i=1}^m 1(s_i^t=j)}\sum_{\{i|s_i^t=j\}}d_i
    \label{eq2}
\end{equation}

If we treat each centroid of cluster $c_j^t$ as a ``query embedding'', Eq. \ref{eq1} can be interpreted as the similarity computation between the document and several queries, determining which of the queries can be answered by the $i$-th token embedding. Thus, the cluster centroid $c_j^t$ plays a similar role as query and we name it ``pseudo query embedding''. Next, the embeddings belong to one cluster compose the new pseudo query embedding by Eq. \ref{eq2}. As the two steps alternatively iterate, the query embeddings that can be answered by the document are explored. Since this process only involves the documents, we can save the embeddings in memory and retrieve them using the real queries which are desired to be resolved.

Since the pseudo query embeddings contain the underlying information of the document that real queries may ask, we use the the pseudo query embeddings as the compressed document embeddings (i.e., the embeddings output by a compressor, as shown in Figure \ref{fig:1}(d)). In the inference stage, we compute the similarity between the pseudo query embeddings $\{c_j\}_{j=1}^k$ and the real query embeddings $\{q_i\}_{i=1}^n$ which can be formulated by the following equations.
\vspace{-1em}
\begin{gather}
    e_q = \texttt{Pooling}(q_1,...,q_n) \label{eq:3}\\
    a_{j} = \texttt{softmax}(e_q\cdot c_j) \label{eq:4}\\
    e_d = \sum_{j=1}^k a_{j}c_j \label{eq:5} \\
    y = e_q\cdot e_d \label{eq:6}
\end{gather}
Eq. \ref{eq:3} means that we pool the query embeddings into a fixed-length embedding $e_q$. Currently, we select the embedding of \texttt{[CLS]} token as $e_q$. As the query is much shorter than the document and usually represents one concrete meaning, we assume this compression will not lose much information. In Eq. \ref{eq:4}, we compute the similarity between the $e_q$ and $c_j$ following a softmax normalization. Then, using the normalized scores as weights, the final document embedding $e_d$ is a weighted sum of the document embeddings, as shown in Eq. \ref{eq:5}. At last, the matching score is computed by the dot product between $e_q$ and $e_d$.

Comparing with existing work, we find that the Poly-Encoder(learnt-k) \cite{poly2019} is equivalent to learning multiple fixed global pseudo query embeddings $\{c_j\}_{j=1}^k$ across all of the documents. That model treats the pseudo query embeddings as learnable parameters which are kept fixed during the inference. It uses the linear combinations of document token embeddings $\{d_i\}_{i=1}^m$ as the compressed document embeddings, taking similarity scores between $\{d_i\}_{i=1}^m$ and $\{c_j\}_{j=1}^k$ as the combination weights. Conversely, the Poly-Encoder(first-k) \cite{poly2019} and ME-BERT\cite{MEBERT2020} use the first $k$ document token embeddings as the pseudo query embeddings, i.e., $\{c_j \}_{j=1}^k=\{d_i\}_{i=1}^k$ and adopt the pseudo query embeddings as compressed document embeddings. 
In contrast to Poly-Encoder(learnt-k), they rely on dynamic pseudo query embeddings. Experimental results on conversation datasets show Poly-Encoder(first-k) is better than the former. 
However, only adopting the first-$k$ document embeddings seems to be a coarse strategy since a lot of information may exist in the latter part of the document. To this end, we present an approach which generates multiple adaptive semantic embeddings for each document by exploring all of the contents in the document. 



\subsection{Large-scale Retrieval Optimization for ANN}

The first-stage retrieval model should calculate the matching scores between the query and all of the documents in the collection. Most existing dense retrieval work adopts Approximate Nearest Neighbor (ANN) searching methods to boost the retrieval process. Faiss\cite{faiss17} is one of the most popular ANN search libraries. It first builds vector index offline and make an ANN vector search based on the index. However, Faiss only supports basic similarity functions like the dot product/Euclidean distance other than the function listed in Eq. \ref{eq:4}-Eq. \ref{eq:6}. To boost in our method using Faiss, we build an index using all of the representations $\{c_j\}_{j=1}^k$ of each document. During inference, we firstly select the $c_j$ which has the highest dot product value with $e_q$ as the final document embedding $e_d$ and compute the matching score using Eq. \ref{eq:6} . Since this operation only involves dot product, it can be accelerated by Faiss. This operation equals to substitute ${a}_{j}$ with $\hat{a}_{j}$ in Eq. \ref{eq:4}.


\begin{equation}
\label{eq:12}
    \hat{a}_{j} = 1(j=\operatorname*{argmax}_{i=1...k}(e_q\cdot c_i))
\end{equation}

As shown in Eq. \ref{eq:12}, we use argmax operation instead of softmax. Such substitution is reasonable since softmax is a derivative and smooth version of argmax \cite{Goodfellow-et-al-2016}. However, only one of the embeddings can pass the argmax function and participate the similarity computation which may impact the retrieval accuracy. To make a trade-off, we firstly recall top-$R$ documents according to Eq. \ref{eq:12} and then calculate accurate scores as described in Eq. \ref{eq:4}-Eq. \ref{eq:6} on the retrieved documents.


\section{Experimental Evaluation}

\subsection{Datasets}

\vspace*{1\baselineskip} 
\noindent{\bf MS MARCO Dataset}\cite{msmarco2016} is a large-scale ad-hoc text retrieval dataset built for two separate tasks: document ranking and passage ranking. These two tasks are adopted in TREC 2019 Deep Learning Track\cite{trec2020} where test sets are provided. The document ranking task contains 3.2 million documents and 0.3 million queries. The passage ranking task contains 8.8 million passages and 0.5 million queries. The main difference between these two tasks exists in the text length, where the average length of the documents and passages are 1124 and 54, respectively. Following most of the existing work, we use MRR to evaluate the development set of MS MARCO and use NDCG to evaluate the TREC test set.

\vspace*{1\baselineskip} 
\noindent{\bf OpenQA Dataset}\cite{DPR2020} is designed for open domain question answering. The authors collect about 21 million documents from Wikipedia as the document collection whose average length is 100. They collect question-answer pairs from several existing QA datasets (e.g., Natural Questions, Trivia QA, SQuAD etc.). Then, they select some documents that contain the answer text and have the highest BM25 scores with the queries, as the positive documents to the query. Currently, the authors release the data of Natural Questions, Trivia QA and SQuAD. For Natural Questions and Trivia QA, the test sets and development sets are available. For SQuAD, only the development set is available. We conduct experiments on this three datasets using top20/100 accuracy as the evaluating metric.

\subsection{Implementation Details}

We initiate the encoder using a BERT base model. Since the BERT base model could handle 512 tokens at most, we truncate each document up to 512 tokens as the input. We set different cluster numbers according to the document length. In the MS MARCO document ranking task, we set the cluster number to 8. In other tasks, we set the cluster number to 4. More experiments about different cluster numbers are shown in the Section 4.5. Since the initial states of the clusters in K-means may influence the performance a lot, we tried two setups: random initiation(i.e., select the hidden states randomly as the initial states) and equal-interval initiation (i.e., cut the documents into equal length intervals and select the cutting locations as the initial states) and find that the equal-interval initiation can outperforms the \textbf{random initiation}. Therefore, we adopt equal-interval initiation in the following experiments. We use AdamW as the optimizer and set the learning rate to 2e-6 and batch-size to 16. During the training, we select one positive document and 4 negative documents for each of the queries. To improve the training efficiency, we adopt the in-batch negatives technique\cite{DPR2020} which takes all other documents in the batch except the positive one as the negative documents for each query. To reduce the discrepancy between the training and inference process, we also adopt the ANCE\cite{ance2020} training paradigm which constructs new hard negative samples using the trained checkpoint of the models. After encoding of the documents, we save them to an IndexFlatIP index provided by Faiss which supports fast inner product calculation. During the inference, we set the number of the documents retrieved by Faiss (i.e., $R$ in Section 3.3) to 1000*$k$.

\subsection{Retrieval Performance}

\vspace*{1\baselineskip} 
\noindent{\bf MS MARCO} Since our goal is to improve the first-stage retrieval performance, we mainly compare our model with other first-stage retrieval models including: docT5Query\cite{doct5query}, DeepCT\cite{DeepCT2019}, RepBERT\cite{RepBERT2020}, ANCE (First-P)\cite{ance2020}, ME-BERT\cite{MEBERT2020}, ColBERT\cite{Colbert2020}. 

\begin{table}[t]
    \small
    \centering
    \begin{tabular}{l|c|c}
    \toprule
         Models & MRR@10 & Recall@1k \\
        \midrule
        DeepCT & 24.3 & 91.0 \\
        docT5Query & 27.7 & 94.7 \\
        RepBERT & 30.4 & 94.3 \\
        ANCE(First-P) & 33.0 & 95.9 \\
        ME-BERT & 33.4 & - \\
        ME-BERT+BM25 & 33.8 & - \\
        ColBERT & 36.0 & 96.8 \\
        Ours & 34.5 & 96.4 \\
    \bottomrule
    \end{tabular}
    \caption{Results on MS MARCO passage ranking dev set.}
    \vspace{-1em}
    \label{tab:1}
\end{table}

Table \ref{tab:1} shows the results on the passage ranking task. We can see that our model outperforms other models except the ColBERT. However, our method is more efficient than ColBERT in terms of the time complexity ($O(mn)$ vs $O(kn)$, $k<<m$). We think the margin is acceptable considering the trade-off between time and accuracy. Comparing to ME-BERT and ANCE, we can see that our proposed method can generate more effective representations. Noticing that ME-BERT adopts a BERT large encoder which has a more powerful language understanding ability than the BERT base encoder in our model, our proposed method is effective enough to bridging the gap.

\begin{table}[t]
    \small
    \centering
    \begin{tabular}{l|c|c}
    \toprule
       Models & MRR@100 & NDCG@10 \\
       \midrule
       ANCE(First-P)  & 37.2* & 61.5 \\
       ME-BERT & 33.2 & - \\
       ME-BERT+BM25 & 34.6 & - \\
       Ours & 39.2 & 62.8 \\
       \bottomrule
    \end{tabular}
    \caption{Results on MS MARCO document ranking dev set(MRR@100) and TREC test set(NDCG@10). The value with * is obtained by the public available code and checkpoint in \url{https://github.com/microsoft/ANCE}}
    \label{tab:2}
    \vspace{-1em}
\end{table}

Table \ref{tab:2} shows the results on the document ranking task. Our model outperforms other models by a large margin. That is probably because the average length of the documents is much longer than the length of passages and our method can make full use of aggregating the semantics of the whole document. 

\begin{table*}[t]
\small
\centering
\begin{tabular}{|l|c|c|c|c|l|l|}
\hline
\multicolumn{1}{|c|}{\multirow{2}{*}{Models}} &
  \multicolumn{2}{c|}{Natural Questions} &
  \multicolumn{2}{c|}{Trivia QA} &
  \multicolumn{2}{c|}{SQuAD} \\ \cline{2-7} 
\multicolumn{1}{|c|}{} &
  Top20 &
  Top100 &
  Top20 &
  Top100 &
  \multicolumn{1}{c|}{Top20} &
  \multicolumn{1}{c|}{Top100} \\ \hline
BM25     & 59.1 & 73.7 & 66.9 & 76.7 & -    & -    \\ \hline
DPR      & 78.4 & 85.4 & 79.4 & 85.0 & 76.4* & 84.8* \\ \hline
BM25+DPR & 76.6 & 83.8 & 79.8 & 84.5 & -    & -    \\ \hline
ANCE     & 81.9 & 87.5 & 80.3 & 85.3 & -    & -    \\ \hline
Ours  & 82.3 & 88.2 & 80.5 & 85.8 & 80.5 & 88.6 \\ \hline
\end{tabular}
\caption{Results on the test sets of Natural Questions and Trivia QA and development set of SQuAD. * indicates the value is obtained by training the model using public code in \url{https://github.com/facebookresearch/DPR}}
\label{tab:3}
\vspace{-1em}
\end{table*}

\vspace*{1\baselineskip} 
\noindent{\bf OpenQA} As for the OpenQA dataset, we compare our model with the DPR model\cite{DPR2020} which is a typical Bi-encoder + independent aggregator structure. Table \ref{tab:3} shows the result of the test set of Natural Questions and Trivia QA and the result of the development set of SQuAD. We can see that our model is better than other models especially in the SQuAD dataset. To explore the possible causal link between the performance and the characteristic of the datasets, we examine the questions corresponding to one document in the training set of different datasets, and find the average number of questions in Trivia QA, Natural Questions and SQuAD are 1.1, 1.4, and 2.7, respectively. It means that the documents in SQuAD corresponds to more questions in comparison with other datasets which may indicate that the passages in SQuAD contain more distinct information than other two datasets. Thus, our method can take full advantage of aggregating different information into clusters.

\subsection{Efficiency Analysis}

\begin{table}[t]
\small
    \centering
    \begin{tabular}{l|c|c}
    \toprule
    Operation & offline & online \\
    \midrule
       Per Document BERT Forward  & 0.9ms & - \\
       Per Document K-means  & 2.1ms & - \\
       Per Document Encoding & 2.3ms & - \\
       Per Query BERT Forward & - & 0.5ms \\
       Retrieval & - & 180ms \\
       Retrieval(w/o optimization) & - & 880ms \\
       Retrieval(independent) & -  &100ms \\
       Retrieval(late interaction) & - & 940ms \\
      \bottomrule
    \end{tabular}
    \caption{Time cost of online and offline computing in MS MARCO document retrieval task.}
    \label{tab:4}
    \vspace{-1em}
\end{table}

We run our model on a single Nvidia Tesla V100 32GB GPU for the MS MARCO document retrieval task and record the time spent by each phase, as shown in Table \ref{tab:4}. Leveraging the powerful parallel computation ability of GPU, the document can be quickly passed through the BERT encoder. It is quite surprising that the K-means algorithm costs more time than BERT given that the time complexity of K-means is less than the deep Transformer in theory. Presumably, this is because our K-means implementation includes a for-loop during the updating step which is not friendly for parallel computing. This part can be optimized using a more parallel friendly implementation. To retrieve documents for new queries, the queries should be firstly encoded. The encoding of queries usually spends less time than the documents because the length is shorter. Next, we record the retrieval time cost by each query with or without the help of the optimization mentioned in Section 3.3. We can find that the optimization can accelerate the retrieval, saving non-trivial time, which confirms the effectiveness of the proposed optimization. To compare our approach with other different aggregators, we also record the retrieval time using independent aggregator and late interaction aggregator. We can see that our model spends an amount of time near to the independent aggregator and outperforms late interaction aggregator by a large margin.

\subsection{Ablation Study}

\begin{table}[t]
\small
    \centering
    \begin{tabular}{l|c}
    \toprule
     Models    &  MRR@100 \\
     \midrule
      random init (k=4)   & 36.8 \\
      w/o ANCE (k=4) & 37.3 \\
      w/o ANCE (k=8) & 37.9 \\
      k=4 & 38.4 \\
      k=8 & 39.2 \\
      k=16 & 39.4 \\
      k=32 & 38.8\\
      \bottomrule
    \end{tabular}
    \caption{Performance of the MS MARCO document ranking dev set under different model settings.}
    \label{tab:5}
\end{table}

We conduct ablation study on the development set of MS MARCO document ranking task. The results are shown in Table \ref{tab:5}. We firstly change the cluster initialization strategy to random. Clearly, the performance drops dramatically since the training becomes unstable. Next, we try to remove the ANCE training mechanism which alleviates the discrepancy between training and inference. We can find that although the performance decreases, it can still outperform the ANCE and the ME-BERT model, showing the effectiveness of the method proposed in this paper. Finally, we compare the performance under different number of clusters ($k=4,8,16,32$). We find that the model achieves the best performance when $k=16$ but the margin leading $k=8$ is not significant. Besides, when $k=32$, the performance drops by a large margin. We infer the reason is that the documents do not have such a number of individual clusters. As a result, the clustering algorithm is hard to converge.

\subsection{How Do the Cluster Centroids Work}

\begin{figure*}
     \centering
     \begin{subfigure}[b]{0.3\textwidth}
         \centering
         \includegraphics[width=\textwidth]{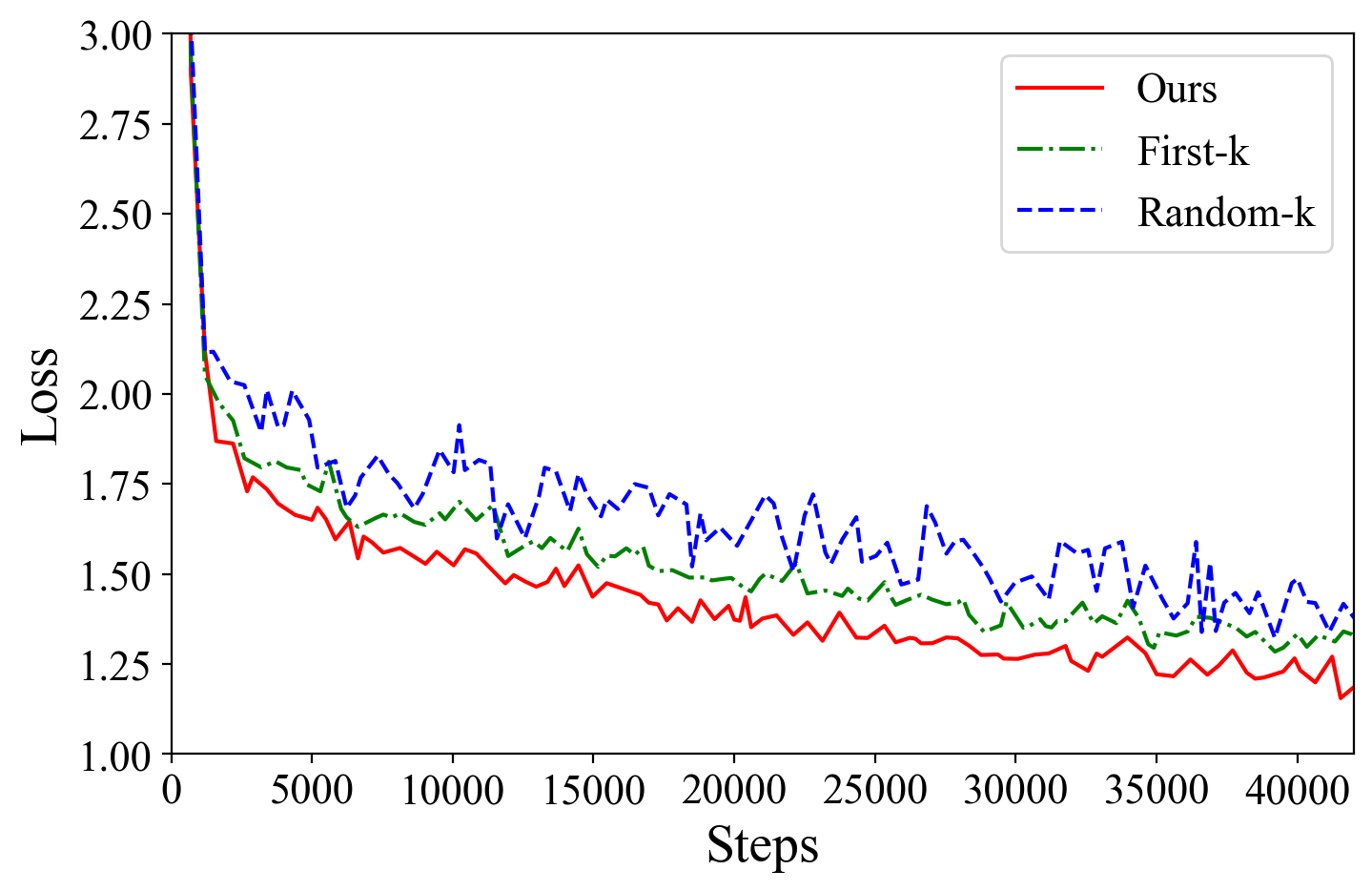}
         \caption{Loss function value.}
         \label{fig:loss}
     \end{subfigure}
     \hfill
     \begin{subfigure}[b]{0.3\textwidth}
         \centering
         \includegraphics[width=\textwidth]{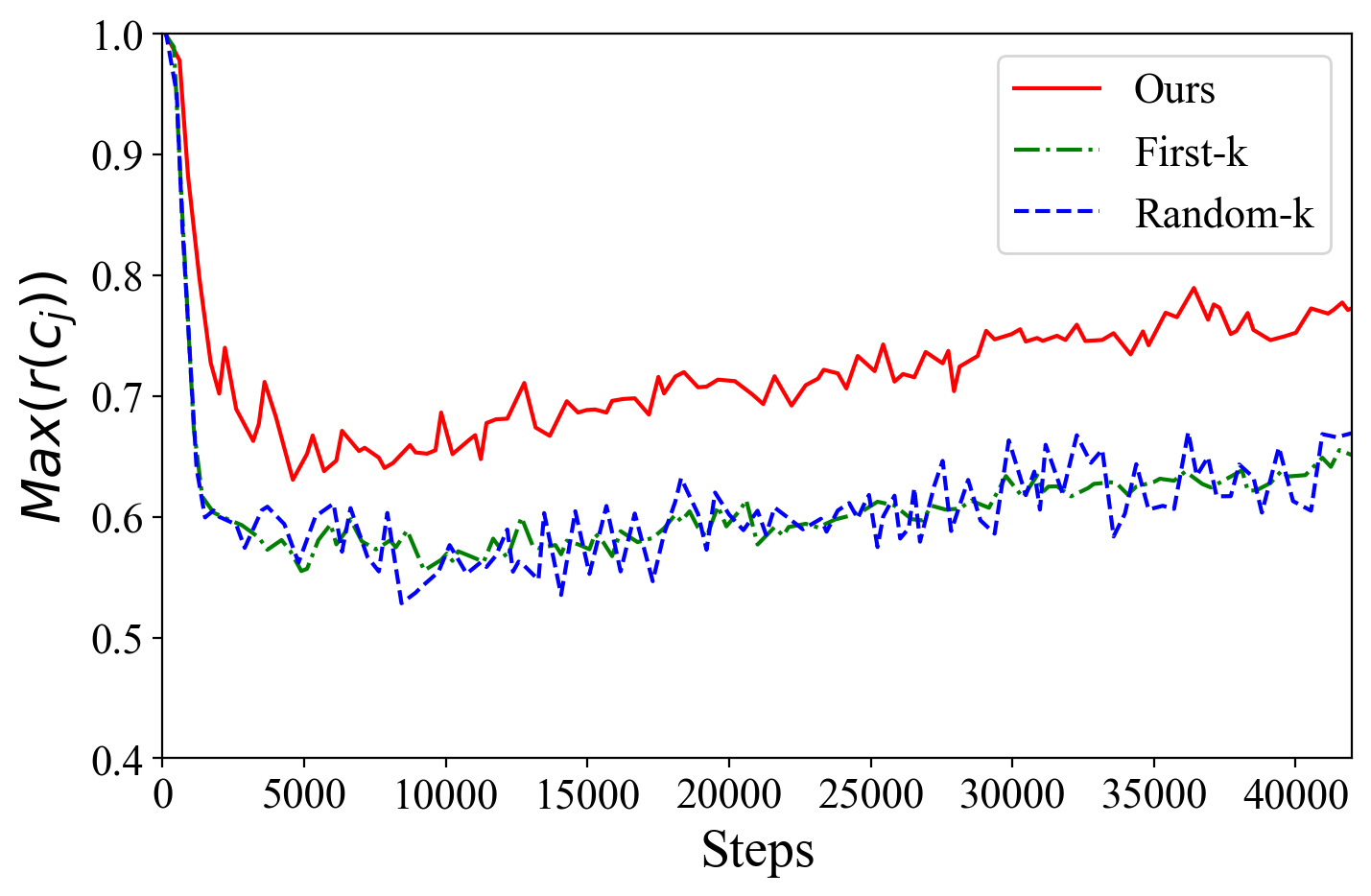}
         \caption{$\text{max}(r(c_j))$.}
         \label{fig:maxcj}
     \end{subfigure}
     \hfill
     \begin{subfigure}[b]{0.3\textwidth}
         \centering
         \includegraphics[width=\textwidth]{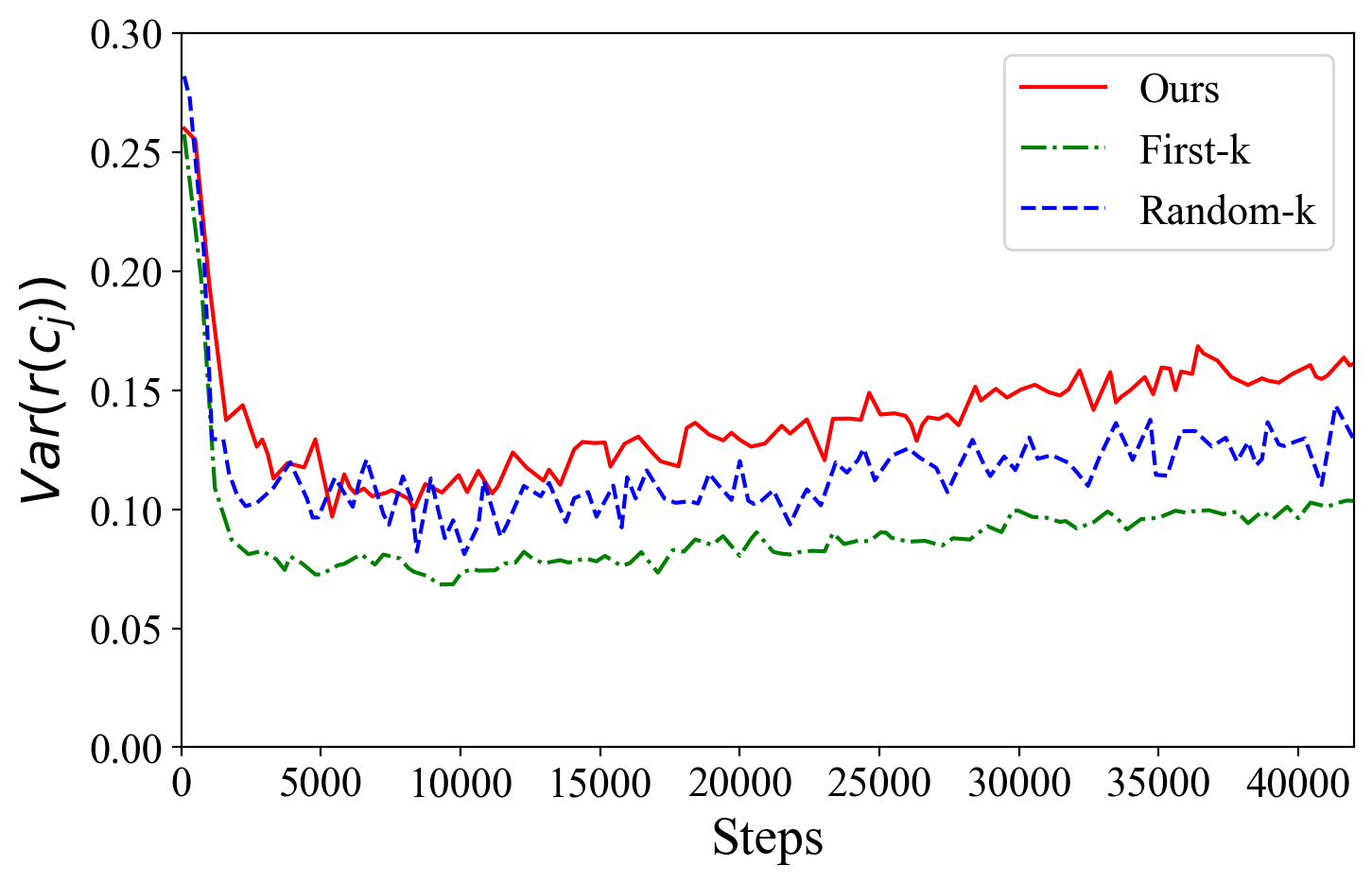}
         \caption{$\text{var}(r(c_j))$.}
         \label{fig:varcj}
     \end{subfigure}

        \caption{Loss, $\text{max}(r(c_j))$ and $\text{var}(r(c_j))$ of different models.}
        \label{fig:values of model}
     \vspace{-0.5em}

\end{figure*}

Although the performance of the ranking metrics like MRR show the effectiveness of the our method, we still need an in-depth view of how the cluster centroid based embeddings improve the model against other methods. In this section, we try to show it by analyzing how the document embeddings affect the value of the loss function.

Given a query $q$ and document collection $D$ where $d\in{D}$ is the relative document of $q$, the training objective is to minimize the loss function in the following form:
\begin{equation}
    \mathcal{L} = -{\log{\frac{\exp(y_d)}{\sum_{t\in D}{\exp(y_t)}}}}
\end{equation}
where $y_d$ and $y_t$ are computed as Eq. \ref{eq:6}. Next, we can see how a single step of gradient descent alters the loss value by analyzing the gradient of the loss function with respect to the document embeddings. For each document embedding $c_j$, we have:
\begin{align}
    \triangledown_d \mathcal{L}_d = & (y_d-1)e_q\triangledown e_d \\
    \triangledown_j e_d = & r(c_j)\triangledown c_j \\
    r(c_j) = & [1 + (\sum_{j'\neq j}a_{j'}(e_q c_j - e_q c_{j'}))]a_j
\end{align}
where $\triangledown_d \mathcal{L}$ means the gradient of loss with respect to document $d$ and $\triangledown_j e_d$ means the gradient of $e_d$ with respect to $c_j$. Details of the derivation are shown in the Appendix. The absolute value of $r(c_j)$ can be interpreted as a weight of how much the $c_j$ can contribute to the loss value. For example, if we feed the model with document embedding producing large positive $r(c_j)$, a single gradient descent step would decrease the loss value faster than small $r(c_j)$.



To verify whether the cluster centroids are more effective than other document embeddings, we compare our model on MS MARCO document ranking task with two other models: the first one adopts the first $k$ token embeddings as the document embeddings like Poly-Encoder(first-$k$)\cite{poly2019} and the second one adopts $k$ randomly selected token embeddings as the document embeddings. Other parts of the model remain unchanged. Ideally, we expect (1) at least one of the document embeddings can match its relative query embedding and (2) multiple document embeddings can capture different semantic information of the document. We use the max value of $r(c_j)$ among multiple document embeddings to evaluate (1) and use the variance of $r(c_j)$ among the multiple embeddings of the same document to evaluate (2). We plot them during the training as shown in Figure \ref{fig:values of model}.

At the beginning of the training, the loss value, $\text{max}(r(c_j))$ and $\text{var}(r(c_j))$ of the models are relatively high and rapidly decrease. When the decreasing of the loss slows down, our model can provide a much higher $\text{max}(r(c_j))$ and lower loss. Besides, $\text{var}(r(c_j))$ of our model is also higher than others indicating the document embeddings are different with each other. We infer that this is because the cluster algorithm expands the distance of the cluster centroids, i.e., $c_j$ and $c_j'$, making the embeddings more distinct with each other. Assuming $i=\text{argmax}_j(r(c_j))$, clustering produces larger $r(c_i)$ and lower $r(c_{i'})$ as shown in Eq. 11. From Eq. 9-10, we can see that large $r(c_i)$ can amplify the impact of $e_q$ to $c_i$ making $c_i$ more approximate to $e_q$. Therefore, the gradient descent can do an accurate update for the specific document embedding $c_i$ towards $e_q$ while leaves $c_i'$ (should represents information other than $e_q$) less changed. As a result, the $c_i$ which is nearer to $e_q$ dominates the loss to reduce more than other models. 


\section{Conclusions}
In this paper, we propose a method to improve the performance of the first-stage retrieval model which is based on Bi-encoder and semi-interactive aggregator. Specifically, our method mimics the real queries by an iterative K-means clustering algorithm. To accelerate the retrieval process, we also optimize the softmax matching function by filtering out some documents using argmax operation. We conduct experiments on the MS MARCO and OpenQA datasets. Through the analysis of the retrieval quality and efficiency, we can confirm the proposed approach is both effective and efficient.



\bibliographystyle{acl_natbib}

\clearpage
\appendix
\section{Appendices}
\label{sec:appendix}

First, the gradient of the loss function with respect to the final document embedding $e_d$ is in the following form:
\begin{align*}
    \triangledown_d\mathcal{L}_d & = -\triangledown\log{\text{softmax}(y_d)} \\
    & = -(\triangledown(e_q e_d) - \triangledown\sum_{d'}y_{d'}(e_q e_{d'})) \\
    & = -e_q \triangledown e_d + y_{d}e_q\triangledown e_d \\
    & = (y_d-1)e_q\triangledown e_d
\end{align*}
where $d'$ includes the positive documents and sampled negative documents during the training. Since we only consider the gradient of the positive document, we ignore the gradients with respect to other documents. Next, ignoring $e_q$ which would not affect the gradient of the document embeddings, we can compute the gradient with respect to the pseudo query embeddings $c_j$ in the following form:
\begin{align*}
    \triangledown e_d = & \triangledown (\sum_{j=1}^k a_j c_j) \\
    = & \sum_{j=1}^k(a_j\triangledown c_j + \triangledown a_j c_j) \\
    = & \sum_{j=1}^k(a_j\triangledown c_j + a_j \triangledown \log{a_j}c_j) \\
    = & \sum_{j=1}^k(a_j\triangledown c_j + a_j[\triangledown(e_q c_j)-\sum_{j'=1}^k a_{j'}\triangledown (e_q c_{j'})]c_j) \\
    = & \sum_{j=1}^k(a_j\triangledown c_j + a_j c_j(e_q\triangledown c_j - \sum_{j'=1}^k a_{j'}\triangledown (e_q c_{j'}))) \\
    = & \sum_{j=1}^k(a_j\triangledown c_j + a_j c_j e_q \triangledown c_j - a_j c_j a_j e_q \triangledown c_j -  \\
    & (\sum_{j'\neq j}^k a_{j'}e_q \triangledown c_{j'})a_j c_j)
\end{align*}

Now, we consider the gradient with respect to a single document embedding $c_j$, we have:
\begin{align*}
    \triangledown_{j}e_d = & [a_j+a_j c_j e_q - a_j c_j a_j e_q - (\sum_{j\neq j'} a_j c_{j'} e_q a_j)]\triangledown{c_j} \\
    = & [a_j+a_j e_q c_j - a_j^2 e_q c_j - a_j e_q (\sum_{j'\neq j} a_{j'}c_{j'})]\triangledown c_j \\
    = & [a_j + a_j(1-a_j)e_q c_j - a_j e_q (\sum_{j' \neq j} a_{j'} c_{j'})]\triangledown c_j \\
    = & [a_j + a_j e_q ((1-a_j)c_j - (\sum_{j'\neq j} a_{j'}c_{j'}))]\triangledown c_j \\
    = & [a_j + a_j e_q (\sum_{j'\neq j}a_{j'} c_j -  (\sum_{j'\neq j} a_{j'}c_{j'}))] \triangledown c_j \\
    = &  [a_j + a_j e_q (\sum_{j'\neq j}a_{j'}(c_j-c_{j'}))] \triangledown c_j \\
    = & [1 + (\sum_{j'\neq j}a_{j'}(e_q c_j - e_q c_{j'}))]a_j \triangledown c_j
\end{align*}

\end{document}